\documentclass[journal=apchd5,manuscript=letter,layout=twocolumn]{achemso}

\usepackage{hyperref}
\setkeys{acs}{keywords = true}
\setkeys{acs}{articletitle = true}
\begin{tocentry}
 \includegraphics[height=3.5cm]{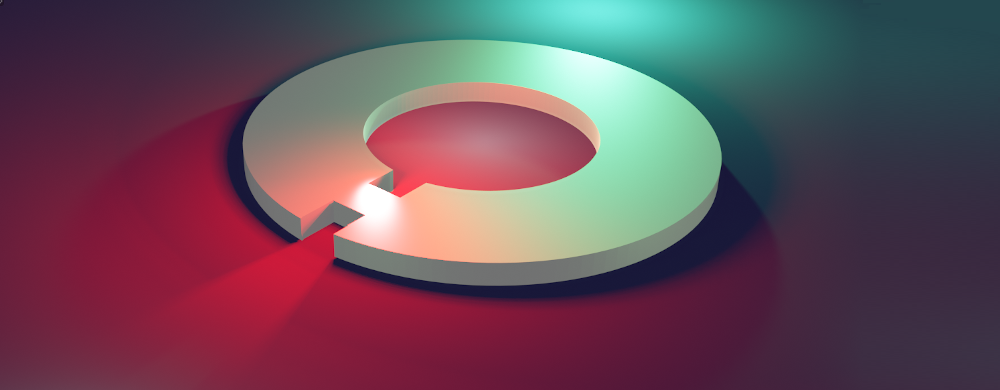}
\end{tocentry}
\usepackage[version=3]{mhchem} 
\usepackage{color}
\usepackage{multirow}
\usepackage{lipsum}
\usepackage{color, soul}
\usepackage{esint}

\newcommand{\Jto}{{\bf J}^{d}}

\newcommand{\rt}{\tilde{\bf r}}
\newcommand{\Ep}{\mathbf{E}}
\newcommand{\rp}{\text{P}}

\newcommand{\FE}{\text{FE}}

\author{Bruno Miranda}
\affiliation{Department of Electrical Engineering and Information Technology, Universit\`{a} degli Studi di Napoli Federico II, via Claudio 21, Naples, 80125, Italy.}
\alsoaffiliation{Institute of Applied Sciences and Intelligent Systems - Unit of Naples, National Research Council, Via P. Castellino 111, Naples, 80131 Italy.}
\author{Vincenzo D'Ambrosio}
\affiliation{Department of Physics {``E. Pancini''}, Universit\`{a} degli Studi di Napoli Federico II, Via Cinthia, I-80126, Naples, Italy.}
\author{Giovanni Miano}
\affiliation{Department of Electrical Engineering and Information Technology, Universit\`{a} degli Studi di Napoli Federico II, via Claudio 21, Naples, 80125, Italy.}
\author{Luca De Stefano}
\affiliation{Institute of Applied Sciences and Intelligent Systems - Unit of Naples, National Research Council, Via P. Castellino 111, Naples, 80131 Italy.}
\author{Carlo Forestiere}
\affiliation{Department of Electrical Engineering and Information Technology, Universit\`{a} degli Studi di Napoli Federico II, via Claudio 21, Naples, 80125, Italy.}
\email{carlo.forestiere@unina.it}
\phone{+39 081 7682007}

\title[An \textsf{achemso} demo]
  {Enhancing Electric Fields \\ in High-Index Resonators \\ by Flux Conservation \\ of the Displacement Current Density}


\begin{document}


\onecolumn
\begin{abstract}
Concentrating light within subwavelength spatial regions is a central topic in nanophotonics. In this letter, we introduce a general	 principle for the subwavelength localization and enhancement of electric fields in high-index resonators, based on the flux conservation of the displacement current density.  We apply this design rule to a ring resonator by locally squeezing its section: since the flux is conserved, the electric field is necessarily enhanced to compensate the reduction of the section area. The introduced principle may constitute an important step toward the control of the displacement current density at the nanoscale,  guiding the design of the topology and the geometry of complex dielectric structures.
\end{abstract}
\twocolumn


Concentrating light within sub-wavelength regions is one of the longstanding goals of nanophotonics \cite{novotny2012principles,schuller2010plasmonics}, and holds promise for enhancing the efficiency of  light emission and photodetection, sensing,  spectroscopy, and heat transfer. This goal may be pursued exploiting the resonance properties of either metallic or high-index nano-objects.

In particular, metallic objects may support plasmonic resonances, which can be described within the electroquasistatic approximation of the Maxwell's equations \cite{Fredkin:03,Bergman:03,Li:03,Wang:06,klimov2014nanoplasmonics} if the resonator is very small compared to the wavelength of operation. High localized electric fields could be then achieved by means of electroquasistatic mechanisms, such as the tip-shape effects in objects with sharp edges or through capacitive coupling in closely spaced metal objects \cite{schuller2010plasmonics}. Although radiative effects can be also exploited to further enhance the local electric field (e.g. \cite{Forestiere:12}), they usually play a minor role.

Unfortunately, the performances of plasmonic devices are typically hampered by metal losses \cite{Khurgin:15}. The need of overcoming them has stimulated the research of alternative platforms, in particular high-index dielectric resonators \cite{garcia2011strong,Evlyukhin:12,kuznetsov2012magnetic,Kuznetsov:16,PhysRevLett.119.243901,jahani2016all}. For instance, silicon resonators, arranged in both periodic \cite{yavas2017chip} and non-periodic\cite{yavas2019unravelling} arrays, have been already proposed as diagnostic platforms for the screening of cancer biomarkers and for the design of chiral sensing platforms \cite{hanifeh2019helicity,Graf:19,mohammadi2019accessible,garcia2019enhanced}.

Contrarily to plasmonic resonances, which have an electrostatic origin \cite{Fredkin:03}, the resonances in high-index dielectric nanoparticles, much smaller than the wavelength of the incident light, can be modeled by the magnetoquasistatic approximation of the Maxwell's equations, where the normal component of the displacement current density vanishes on the surface of the object \cite{Forestiere2020}. The existence of such a profound difference between these two types of resonators suggests that electrostatic-inspired strategies, which are very effective in metal resonators, may not be as good in enhancing local fields in dielectric ones.

Several scientific contributions have already investigated the field enhancement and localization in high/moderate-index resonators\cite{garcia2011strong,luk2017hybrid,wang2019broken,Mignuzzi:19}, pointing out that electric field  {\it hot spots} are typically localized in the interior of the resonator \cite{kapitanova2017giant}, in sharp contrast with plasmonic ones. Although, this is certainly an advantage for non-linear applications \cite{grinblat2016enhanced,Ghirardini:18,Carletti:18,koshelev2020subwavelength}, it constitutes a limitation when the goal is to enhance the emission from nearby molecules or quantum emitters. Aware of this limitation, in Ref. \cite{Yang:18} , the authors introduced a clever strategy to generate an {\it external} hotspot making a slot aperture in a high-index dielectric disk, achieving very significant enhancement of the electric field, based on the action of the so called {\it toroidal} dipolar mode (i.e. $\text{HEM}_{12\delta}$) of a disk. Another very effective strategy was proposed also in Ref. \cite{robinson2005ultrasmall} where the authors exploited the jump of the {\it normal} component of the electric field at the interface between a high-index dielectric material and the air.

\begin{figure}[ht!]
    \centering
    \includegraphics[width=0.7\columnwidth]{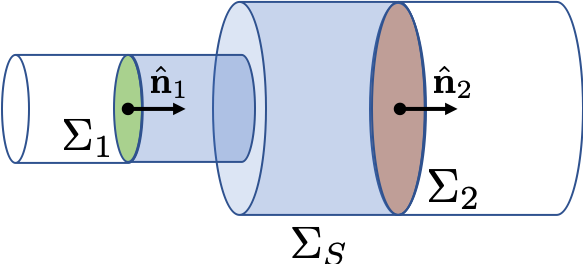}
    \caption{Sketch of a portion of the resonator having varying cross section. $\Sigma_S$ is a portion of the physical boundary of the resonator, $\Sigma_1$ and $\Sigma_2$ are purely mathematical surfaces inside the resonator.}
    \label{fig:Sketch}
\end{figure}

Given the complementary nature of plasmonic and dielectric resonances, the problem of achieving high field enhancement in high-index objects requires renovated methods, which can be inspired by inherent properties of magnetoquasistatic resonances. {\it Engheta et al.} \cite{engheta2007circuits} already envisioned that high-index materials can play the role of nanoscale optical {\it conduits} for the displacement current density field, analogously to the role that wire conductors play for direct currents in electric circuits. 

In this paper, we further carry on this analogy, introducing a radically new strategy to achieve high electric field enhancement both in the inside and outside of a high-index resonator. Our strategy is based on the flux conservation of the displacement current density field. In fact, as a reduction of the section of a wire conductor entails a local increase of the current density; analogously, a reduction in the section of the high-index material is connected with an increase of the displacement current density and therefore of the electric field. We demonstrate this general principle by  squeezing the section of a ring resonator, where the fundamental mode is a displacement current density loop associated to a magnetic dipole. Because the flux of the displacement current is conserved, the magnitude of the displacement current density has to increase to compensate the decrease of the section. As a result, the electric fields are also enhanced outside of the resonator, thanks to the continuity of the tangential component of the electric field. 

\begin{figure}
\centering
\includegraphics[width=\columnwidth]{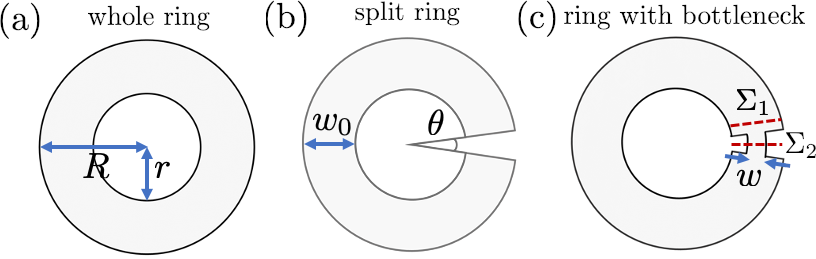}
\caption{Top view of the three investigated ring resonators:  whole ring (a),  split ring (b), and ring with a bottleneck (c), with minor radius $r$, major radius $R=2 r$, and height $h$. The width of the ring is $w_0=r$. In the split ring, we removed  an angular sector of aperture $\theta$. In (c), the bottleneck has width $w$.}
\label{fig:schematics}
\end{figure}

We apply this principle to high- and moderate- index resonators under different excitation conditions, showing strong localization and enhancement of the electric field, with potential applications in surface-enhanced molecular spectroscopies \cite{li2013surface} or single-photon emission enhancement \cite{tapar2020enhancement}, surface-enhanced Raman spectroscopy \cite{caldarola2015non}, single-molecular detection and biosensing \cite{pryce2011compliant,bontempi2017highly}. 


\subsubsection{Flux Conservation in High Index Resonators}

To demonstrate the flux-conservation principle, let us consider a  homogeneous, isotropic, non-magnetic linear material having relative permittivity $\varepsilon_R$. It occupies a bounded domain $\Omega$, with a regular boundary $\partial \Omega$. We define its characteristic size $l_c$ as a chosen linear dimension of the object and its size parameter $x = 2 \pi l_c /\lambda$, where $\lambda$ is the vacuum wavelength. Here, we briefly recall some of the properties of magnetoquasistatic resonances of high-index dielectric objects, introduced in Ref. \cite{Forestiere2020}. These are connected to the eigenvalues of a magnetostatic integral operator $\mathcal{L}_m$ that gives the vector potential as a function of the displacement current density field $\Jto$:
\begin{equation}
y_n^2 \, \mathcal{L}_m \left\{ \Jto_h \right\} \left( \rt  \right) =  \Jto_h \left( \rt  \right) \qquad \forall \rt  \in  \Omega,
\label{eq:MQSproblem}
\end{equation}
with the condition
\begin{equation}
    \Jto_h \left( \rt  \right) \cdot \hat{\mathbf{n}} \left( \rt  \right) = 0 \qquad \forall \mathbf{r} \in \partial \Omega,
    \label{eq:NormalCurrent}
\end{equation}
where
\begin{equation}
\mathcal{L}_m \left\{ \Jto \right\} \left( \rt \right)  = \iiint_{\tilde{\Omega}} \frac{\Jto \left( \rt' \right)}{\left| \rt - \rt' \right|}  d^3 \rt',
\label{eq:MQSoperator}
\end{equation}
$\rt = \mathbf{r}/l_c$, $\tilde{\Omega}$ is the scaled domain, $\partial \tilde{\Omega}$ is the boundary of $\tilde{\Omega}$, $\hat{\bf n}$ is normal to $\partial \tilde{\Omega}$ pointing in the vacuum region. Equation \ref{eq:MQSproblem} holds in the weak form in the functional space constituted by the functions which are div-free within $\Omega$ and having zero normal component and equipped with the inner product
\begin{equation}
    \langle \mathbf{A}, \mathbf{B} \rangle_\Omega  = \iiint_\Omega \mathbf{A} \cdot \mathbf{B} \, d^3 \rt .
\end{equation}

Due to Eq. \ref{eq:NormalCurrent} the magnetoquasistatic oscillations associated to the current density eigenmodes $\Jto_h$ arise from the interplay between the polarization energy stored in the dielectric and the energy stored in the magnetic field. The spectrum of the magnetoquasistatic operator $\mathcal{L}_m$ is discrete \cite{Forestiere2020,forestiere2020resonance}. The resonance of the current mode $\Jto_h$  is characterized by a real and positive eigenvalue $y_n$, which is size-independent. The induced current density fields $ \Jto_h$ have a non-zero curl within the particle, but are div-free and have a vanishing normal component on the particle surface.

The resonance angular frequency $\omega_h$ of the h-th resonance is given by:
\begin{equation}
    \omega_h = \frac{c}{l_c \sqrt{\varepsilon_R}} y_h.
\end{equation}

The resonant electric fields $\mathbf{E}_h$ are immediately connected to the eigenvalue $y_n$ and to the displacement current density mode $\Jto_h$ through the relation:
\begin{equation}
    \mathbf{E}_h = \frac{\mu_0 l_c^2}{y_n^2} \, \Jto_h.
    \label{eq:EigenField}
\end{equation}

Due to Eq. \ref{eq:NormalCurrent}, the normal component of the displacement current density field vanishes on the surface of the object: this immediately implies that the flux of any displacement current density field mode through any portion of the boundary of a high-index resonator vanishes. 

In the following, we turn this property into a strategy for enhancing the electric field in high-index resonators. Let us consider a closed surface $\Sigma$, which is the union of three sections $\Sigma_1$, $\Sigma_2$, $\Sigma_S$  such that
$\Sigma = \Sigma_1 \cup \Sigma_2 \cup \Sigma_S $, as sketched in Fig.\ref{fig:Sketch}. $\Sigma_S$ is a portion of the physical resonator boundary $ \Sigma_S \subset \partial \Omega$, where the normal component of the current density vanishes, while  $\Sigma_1$ and $\Sigma_2$ are purely mathematical surfaces contained in the interior of $\Omega$, with normal unit vector $\hat{\mathbf{n}}_1$ and $\hat{\mathbf{n}}_2$ and surface areas $A_1$ and $A_2$. Therefore, we have that the flux of the displacement current is conserved
\begin{equation}
  \iint_{\Sigma_1} \Jto_h \cdot \hat{\bf{n}}_1 \, d\tilde{S}'  = \iint_{\Sigma_2} \Jto_h \cdot \hat{\bf{n}}_2 \, d\tilde{S}' 
    \label{eq:FluxConservationI}
\end{equation}
Eq. \ref{eq:FluxConservationI} holds for any mode $h$ of any high-index resonator. We now make the additional assumption that there exists a density current mode, e.g. $\Jto_h \left( \mathbf{r} \right)$, which is directed along $\hat{\mathbf{n}}_1$  and $\hat{\mathbf{n}}_2$ on the surfaces $\Sigma_1$ and $\Sigma_2$, respectively
\begin{equation}
 \Jto_h \left( \mathbf{r} \right)  = J^d_{h, i} \left( \mathbf{r} \right) \hat{\bf n}_i \qquad \mathbf{r} \in \Sigma_i  \qquad i =1,2
\label{eq:Hyp}
\end{equation}
and we also assume that the corresponding flux is non-vanishing. In the next section, we show that a high-index ring resonator exhibits topologically protected modes satisfying these hypotheses. By using Eq.~\ref{eq:Hyp} in Eq.~\ref{eq:FluxConservationI} we obtain:
\begin{equation}
   \frac{ \langle J^d_{h,1} \rangle_{\Sigma_1}}{  \langle J^d_{h,2} \rangle_{\Sigma_2}} = \frac{A_2}{A_1} 
    \label{eq:FluxConservationII}
\end{equation}
where $\langle J^d_{h,i} \rangle_{\Sigma_1}$ is the average of $J^d_{h,i}$ on $\Sigma_i$. By combining Eqs.~\ref{eq:EigenField} with Eq.~\ref{eq:FluxConservationII}, we directly obtain that
\begin{equation}
   \frac{ \langle E_{h,1}\rangle_{\Sigma_1}}{\langle E_{h,2} \rangle_{\Sigma_2}} =  \frac{A_2}{A_1} 
   \label{eq:FluxConservation}
\end{equation}
where $\langle E_{h, i} \rangle_{\Sigma_i}$ is the average of the electric field on $\Sigma_i$.
Equation~\ref{eq:FluxConservation} is the central result of this paper and constitutes a general principle to achieve high field enhancement in high-index resonators. By squeezing the section of the resonator, as schematized in Fig.~\ref{fig:schematics}c, the average electric field of the mode increases to keep the flux constant, counterbalancing the reduction of the surface area. 

\label{sec:Results}
\subsection{Magnetoquasistatic modes} 
\begin{figure}[ht!]
    \centering
    \includegraphics[width=7cm]{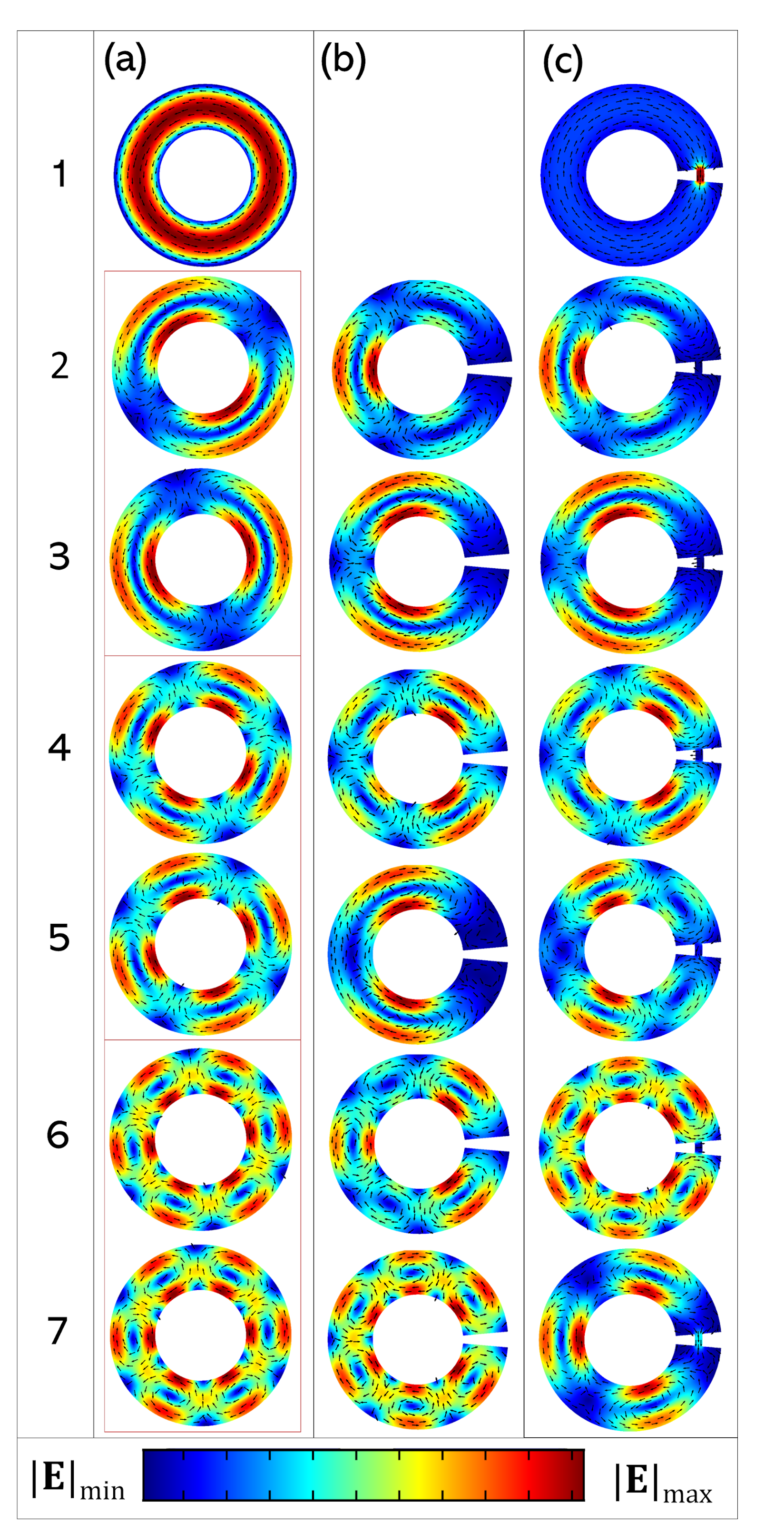}
    \caption{Magnetoquasistatic displacement current density modes of ring resonators with minor radius $r$, major radius $R=2r$ and height $h \ll R$. Whole ring (a), split ring (b), and ring with a bottleneck of width $w = w_0/6$  (c). Pairs of degenerate modes are enclosed in red boxes.}
    \label{fig:modal_analysis}
\end{figure}{}
Looking for a mode that could be a suitable candidate for the application of the flux-conservation design principle, we investigate the magnetoquasistatic modes of a dielectric ring.  Dielectric ring resonators have been recently studied experimentally by {\it Zenin et al.} to control scattering directionality \cite{zenin:20}. Here, we investigate the ring sketched in Fig.~\ref{fig:schematics}a with minor radius $r$, major radius $R$, with $R=2r$, corresponding width $w_0=R-r=r$, and a thickness $h$. 
 To simplify our analysis, and only within the current subsection, we make the additional assumption that the ring has thickness $h$ much smaller than $R$: under this hypothesis we can use the approximated method  introduced in Ref. \cite{forestiere2019electromagnetic} to calculate the magnetoquasistatic modes of the ring, instead of the three-dimensional but more computationally intensive method presented in \cite{Forestiere2020}. 
 
 \begin{figure*}[ht!]
    \centering
    \includegraphics[width=0.8\textwidth]{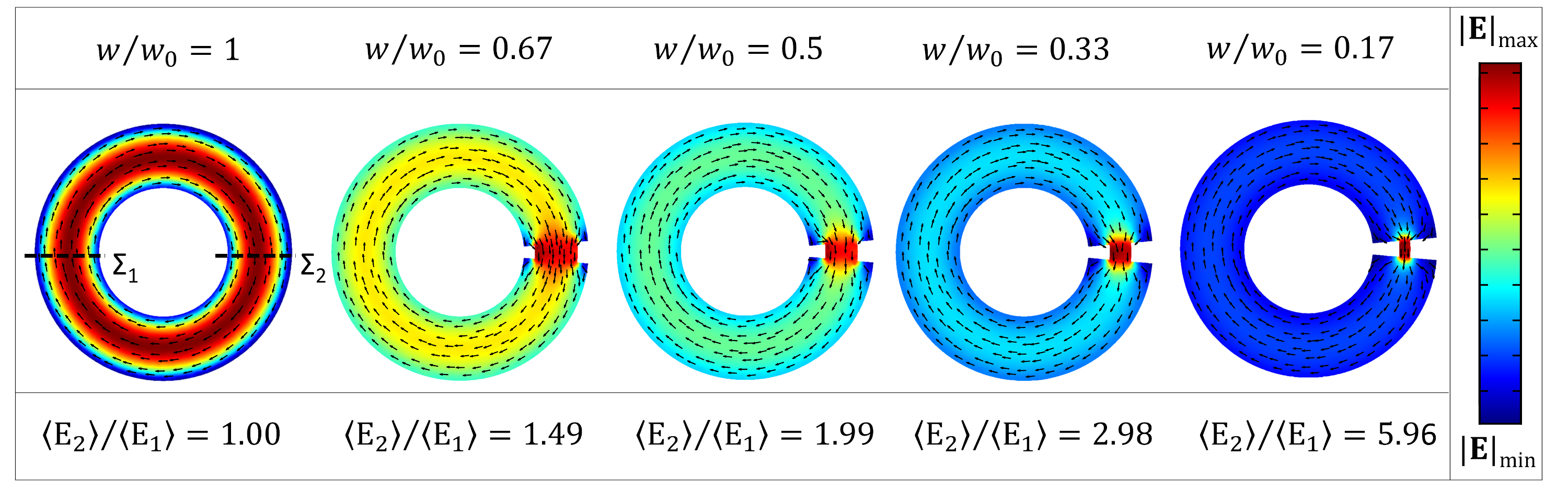}
    \caption{Fundamental magnetoquasistatic mode of ring a resonator with a bottleneck of varying width $w$. Field localization is achieved by decreasing $w/w_{0}$.}
    \label{fig:field_localization}
\end{figure*}{}
The first seven modes exhibited by the ring resonator are shown in Figure~\ref{fig:modal_analysis}a, which are are independent of $h$ as long as $h \ll l_c$.
The fundamental mode $\Jto_1$ is a displacement current loop circulating along the toroidal direction, shown at the top of Fig.~\ref{fig:modal_analysis}a. The next modes exhibit two or more counter-rotating loops of current density field, corresponding to counter-directed magnetic dipole moments. 

The  mode $\Jto_1$  fulfills all the requirements needed of the flux-conservation design principle since i) the current density is always directed along the toroidal direction and ii) the associated flux is non-vanishing. In particular, on the two sections $\Sigma_1$ and $\Sigma_2$ the displacement current density is directed along the corresponding normal unit vector $\hat{\mathbf{n}}_1$ and $\hat{\mathbf{n}}_2$, respectively. Instead, the high-order modes $\Jto_{2-8}$ shown in Fig.~\ref{fig:modal_analysis}a exhibit a vanishing flux at any section of the ring, and  are not useful for our goals.

Thus, we introduce a bottleneck of width \textit{w} in the ring in correspondence of the angular sector of aperture $\Delta \theta = 10^\circ $, as shown in Fig. \ref{fig:schematics}b, investigating in  Fig. \ref{fig:field_localization} the change in the fundamental magnetoquasistatic mode as the bottleneck reduces. As long as the width \textit{w} of the bottleneck remains finite, the topology of the domain preserves the existence of the current-loop mode. Moreover, as the width $\textit{w}$ is reduced, the amplitude of the displacement current density is enhanced within the bottleneck to keep the flux of $\Jto_1$ constant.

\begin{figure}
    \centering
    \includegraphics[width=\columnwidth]{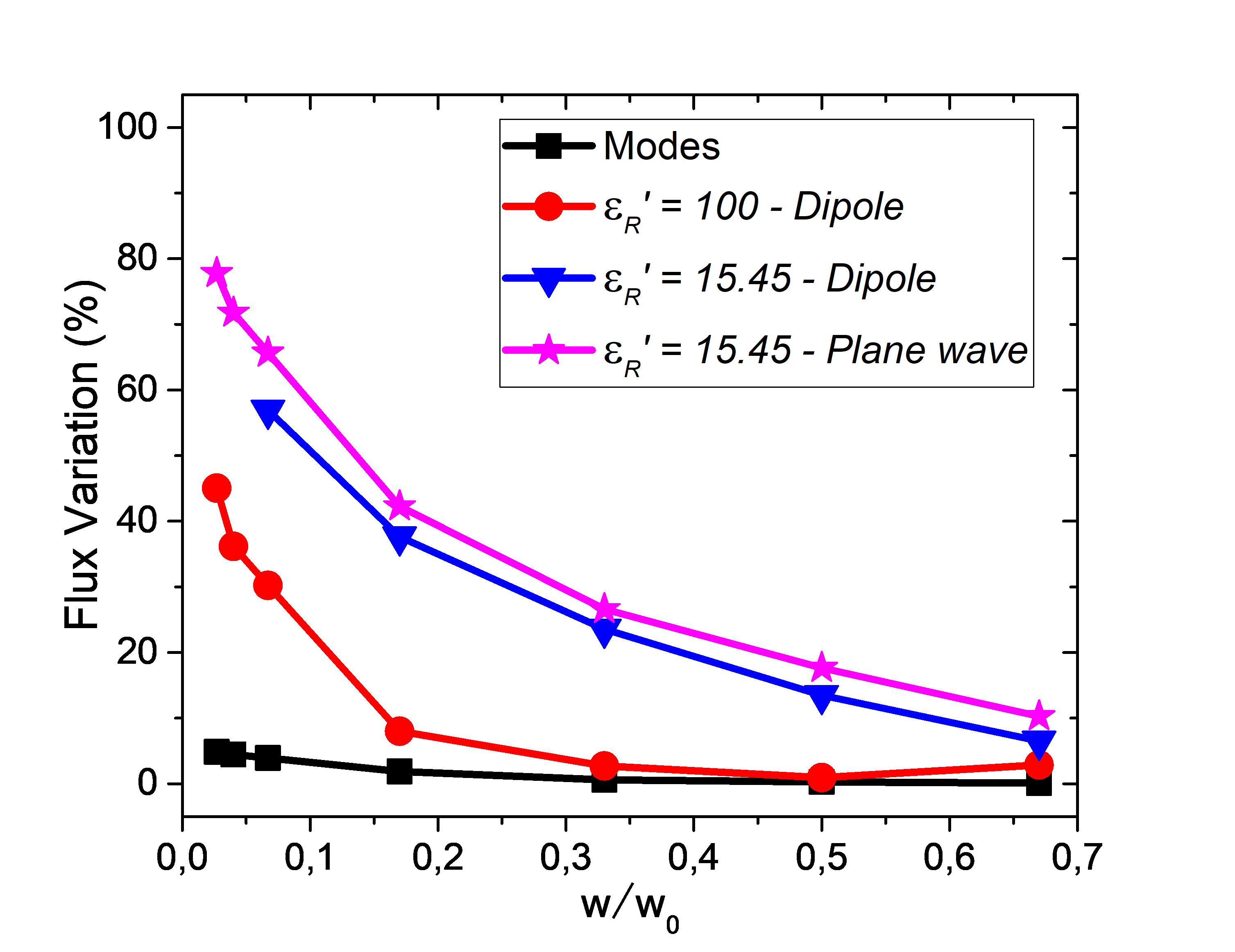}
    \caption{Relative variation of the flux (\%) between the two cross sections $\Sigma_1$ and $\Sigma_2$ of the ring resonator defined in Fig. \ref{fig:schematics}c, as a function of the bottleneck relative width $w/w_0$. The analysis has been performed for the fundamental magnetoquasistatic mode of a ring  of Fig. \ref{fig:field_localization} (black squares), for the electric field {\it state} at the low frequency peak of the high index resonator of Fig. \ref{fig:Hindex} (red circles), and of the silicon resonator of Figs.  \ref{fig:Si} and Fig. S5, excited by an electric point dipole (blue triangles) and by a plane wave polarized in the plane of the ring (purple stars).}
    \label{fig:FluxVariation}
\end{figure}{}

To support this claim, in Fig.~\ref{fig:FluxVariation} we show with a black line the relative variation of the flux of the displacement current density $\Jto_1$ between the two sections $\Sigma_1$ and $\Sigma_2$, defined in Fig.~\ref{fig:schematics}, as a function of $w/w_0$. In the investigated $w/w_0$ range, the flux variation, $\left| \Phi_2 - \Phi_1 \right| / \Phi_1$ is lower than $5\%$, confirming our design principle.

Figure \ref{fig:modal_analysis}c shows the first magnetoquasistatic modes of the ring with the bottleneck of width $w=w_0/6$. The higher-order modes $\Jto_{2-8}$ are only slightly affected by the introduction of the bottleneck. Specifically, the symmetry breaking lifts the degeneracy of the  modes  $\Jto_{2-3}$, $\Jto_{4-5}$, $\Jto_{6-7}$ of the whole ring, but since the net-flux of the displacement current through any section of the ring is zero, we do not observe any field localization.

To further underline the role played by the topology of the resonator geometry, we also show 
in Figure~\ref{fig:modal_analysis}b the magnetoquasistatic modes of a split ring resonator. The displacement current density loop embracing the whole structure does not exist in this case, due to the different topology (genus) of the object. The remaining modes are similar to the corresponding modes of a whole ring, and thus are less affected by the change in the topology of the structure.

\subsection{Full-wave analysis}
So far, we have investigated how to apply the flux-conservation design principle to the magnetoquasistatic modes of the ring resonators, which were inherently independent of the excitation conditions. In this section, guided by the previous modal analysis, we turn to the solution of the full-wave 3D scattering problem in the presence of an external excitation using finite-element-method simulations, carried out in COMSOL Multiphysics.

\begin{figure*}
    \centering
    \includegraphics[width=0.9\textwidth]{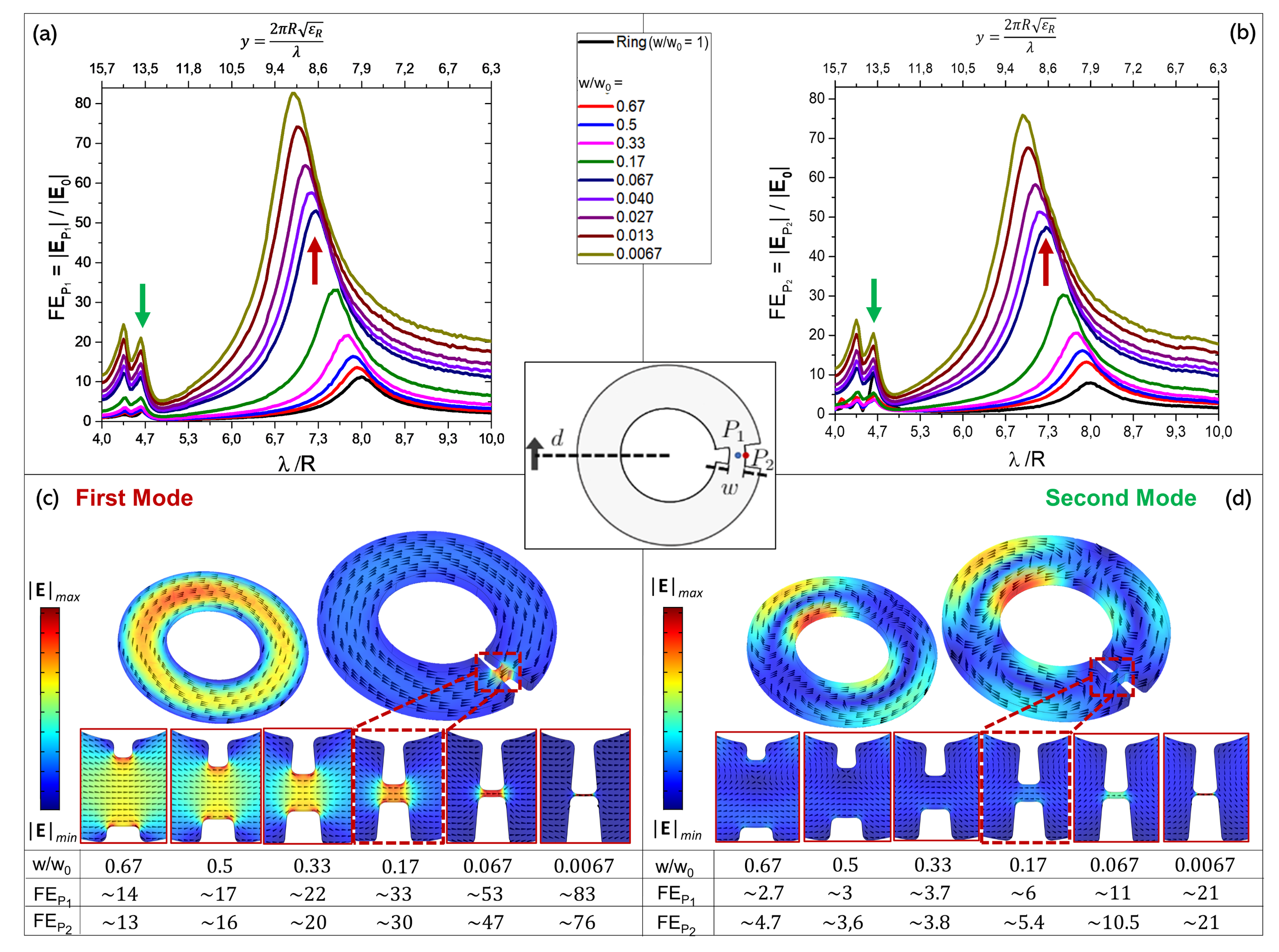}
    \caption{High index ($\varepsilon_R = 100 - 2 i $)  ring resonator with major radius $R$, minor radius $r=R/2$, thickness $h=R/10$, and bottleneck relative widths $w/w_{0}$, excited by the dipole shown in the inset with $d=3r$. FE as a function of the free-space wavelength, probed at the points in the dielectric region $\text{P1}$ (a), and in vacuum  $\text{P2}$ (b). The plots also include the FE of the whole ring. Electric field distributions at the fundamental mode resonant peak, indicated by the red arrow, (c) and at the peak of the second resonant mode, indicated by the green arrow, (d) for different $w/w_{0}$ values. Zoomed-in view of the bottleneck regions and a table reporting the FE values at $\text{P}_1$ and $\text{P}_2$, corresponding to the different ratios $w/w_{0}$, inside and outside the structures.}
    \label{fig:Hindex}
\end{figure*}


It is important to point out that a plane wave, linearly polarized in the plane of the ring and propagating orthogonally to it, is not able to excite the fundamental loop of the displacement current density. Nevertheless, such a mode can be excited by a plane wave propagating within the plane of the ring, or an electric dipole located in proximity of the ring. A dipolar excitation may model  an emitting molecule, a quantum dot, or metal nanoparticle located nearby. In the following, we consider a dipole excitation, while the analysis carried out for a plane wave excitation has been reported in the SI.

In Figure~\ref{fig:Hindex}, we investigate high-index ring resonators with high relative permittivity $\varepsilon_R = 100 - 2 i $. The ring resonator has rectangular cross section, with minor radius $r$, major radius $R=2r$ and thickness $h=R/10$, centered in the origin of a Cartesian coordinate system with its flat faces parallel to the $xy$ plane.  In all the investigated cases, to improve numerical convergence, we introduced a smoothing of the bottleneck corners (fillet) with curvature radius of $r/15$. 

We investigate the field enhancement by varying the  bottleneck width $w$. The total electric field $\Ep \left( \text{P} \right)$ is probed at the two different points shown in the inset: the point $\text{P}_1$ is located within the dielectric region in the middle of the bottleneck $\text{P}_1 = \left( r+\frac{w_0}{2}, 0, 0 \right)$, the point $P_2$ is located in the air just outside the boundary of the resonator $ \text{P}_2 = \left( r + \frac{w_{0}+w}{2}, 0, 0 \right)$. In the same two points, we also calculate the electric field $\mathbf{E}_0 \left( \text{P} \right)$ generated by the exciting dipole, when it is free-standing, namely in the absence of the dielectric object.
The electric field enhancement $\FE$ is defined as the ratio of these two quantities:
\begin{equation}
    \text{FE} \left( \rp,  \lambda \right) = \frac{\left\| \Ep \left( \rp,  \lambda \right) \right\|}{\left\| \mathbf{E}_0 \left( \rp,  \lambda \right) \right\|}
    \label{eq:FE}
\end{equation}

\begin{figure*}[ht!]
    \centering
    \includegraphics[width=0.9\textwidth]{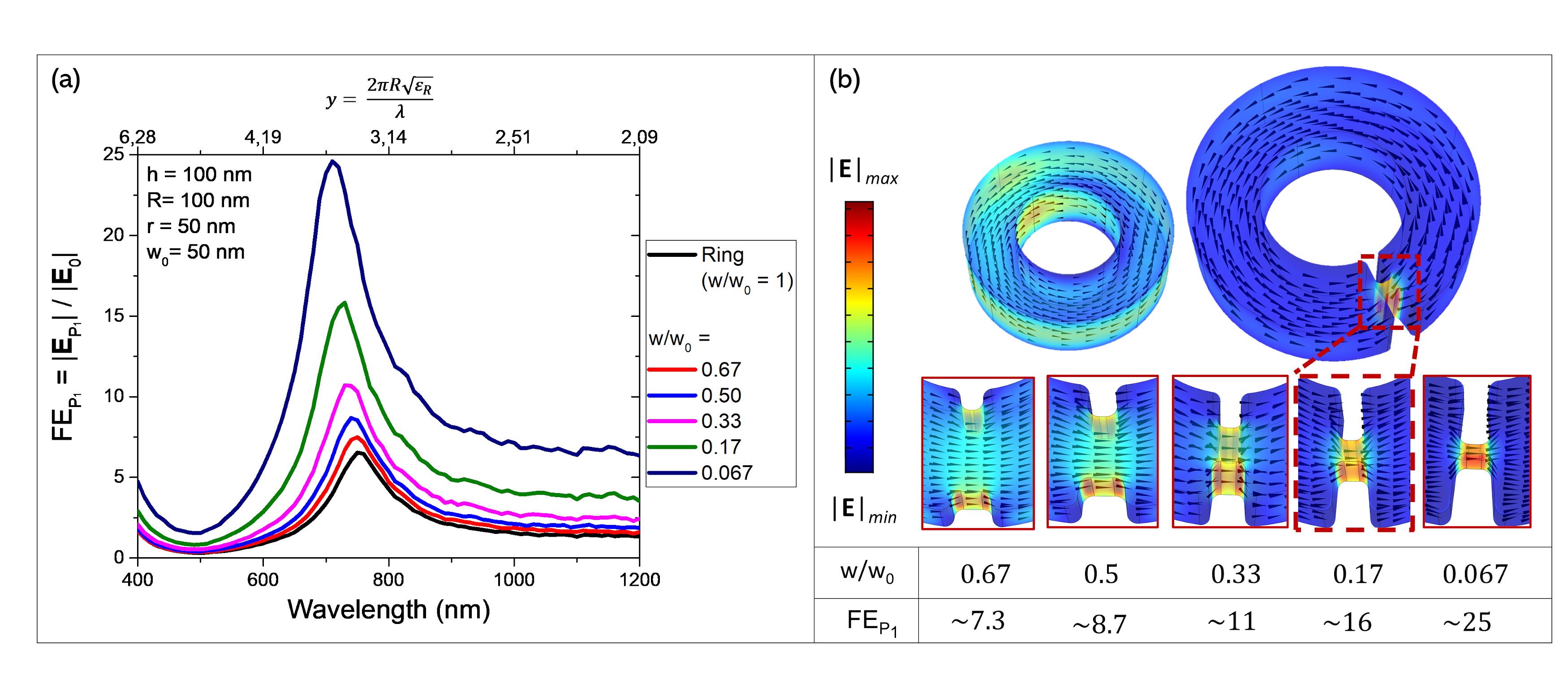}
    \caption{Moderate index ($\varepsilon_R = 15.45-0.1456i$)  ring resonator with major radius $R=100$ nm, minor radius $r=R/2$, thickness $h=100$ nm, and different bottleneck widths $w/w_{0}$, excited by a dipole as exemplified in the inset of Fig. \ref{fig:Hindex} with $d=3R$. (a) Field enhancement as a function of the free-space wavelength, probed in the dielectric region at the point $\text{P}_1$. (b) Electric field distribution at the resonant peak of the fundamental mode for different bottleneck width ratios $w/w_0$ with a zoomed-in view of the bottleneck regions.}
    \label{fig:Si}
\end{figure*}{}

We show the field enhancement as a function of the free-space wavelength for several relative bottleneck widths $w/w_0$ at $\text{P}_1$ (Fig.~\ref{fig:Hindex}a) and at $\text{P}_2$ (Fig.~\ref{fig:Hindex}b). In both cases, the FE spectrum of the whole ring resonator ($w=w_0$) is also shown for comparison.  Since the electric fields of the fundamental mode is tangent to the boundary of the ring, and the tangential component of the electric field are always conserved, the electric field enhancement values in in Fig.~\ref{fig:Hindex}b are  roughly the same as the values in Fig.~\ref{fig:Hindex}a. As we reduce the bottleneck width $w$, the peak value of field enhancement increases from $\displaystyle\max_\lambda \left\{\FE \left( \text{P}_1, \lambda \right) \right\}=11$ for $w=w_0$, to $\displaystyle\max_\lambda \left\{\FE \left( \text{P}_1, \lambda \right) \right\}=83$  for ${w}/{w_0}=0.0067$. 
The electric field localization within the bottleneck is shown in Figure~\ref{fig:Hindex}c
in correspondence of several ratios $w/w_{0}$. 

To quantify the validity of the flux-conservation principle, we show in Fig.~\ref{fig:FluxVariation} the variation of the flux of the electric field between the cross sections $\Sigma_1$ and $\Sigma_2$ of the ring, as defined in Fig. \ref{fig:schematics}c. Differently from what was done in the previous section, here the analysis is performed on the {\it state} of the electric field, and not on the modes. As long as $w \ge 0.15 w_0$ the flux undergoes a variation of less than $10 \% $. Significant violations of the flux conservation happen only for values of $w$ less than  $ 0.1 w_0$.

In Fig. S2 of the Supporting Information, we also investigate the spatial decay of the field enhancement in the external region (air) as a function of the distance from the resonator's walls, concluding that the field enhancement supported by this platform are easily accessible outside the resonator.

The second peak at higher frequencies in  Figure~\ref{fig:Hindex}a-b is caused by the second mode of the ring with bottleneck, shown in Fig.~\ref{fig:modal_analysis}c. This mode exhibits vanishing flux at any section of the ring, therefore flux-conservation does not imply field enhancement, which in facts is significantly lower compared to the first peak. Consistently, the field distributions do not exhibit localization within the bottleneck, as shown in Figure~\ref{fig:Hindex}d. Appreciable FE values at the second peak are only obtained when the ratio $w/w_{0}$ reaches very low values (less than $0.067$), due to a spectral overlap with the fundamental mode. 

However, although permittivities of values 100 and higher are routinely available at microwaves frequencies for several applications including filters, resonators and antennas\cite{richtmyer1939dielectric,kajfez1998dielectric,Long:83,Mongia:94}, in the visible or near infrared only materials with moderate permittivities, such as Silicon or GaAs  are currently available. In the following, we consider silicon rings, with $\varepsilon_R = 15.45-0.1456i$. The ring resonator has major radius $R=100$ nm, minor radius $r=R/2$, thickness $h=100$ nm, and different bottleneck widths $w/w_{0}$. In this case, the hypotheses behind the magnetoquasistatic approximation are not fully verified.

It is immediately apparent that the values of field enhancement reached by introducing a bottleneck in the silicon ring are lower if compared to high-index materials, as shown in figure~\ref{fig:Si}a. For instance, the field enhancement $\FE$ obtained for $w/w_0 = 0.067 $ is about $25 \times$, compared to about $50\times$ achieved for the investigated high-index resonator. This is because, for objects of moderate permittivities, the hypotheses behind the magnetoquasistatic approximation are not fully verified, and a non-vanishing normal component of the displacement current density  to the boundary of the particle arises, causing a dispersion of the flux  and a consequent reduction of the maximum achievable field enhancement. This is confirmed by looking at the variation of the flux between the sections $\Sigma_1$ and $\Sigma_2$ of the ring  in Fig. \ref{fig:FluxVariation}. For a bottleneck of width $w = 0.2 w_0 = 10nm $, about $40\%$ of the flux is dispersed. Comparable values of field enhancement  and similar considerations for the flux dispersion also hold for a silicon ring excited by a plane wave propagating in the plane of the ring, investigated in Fig. S5 of the Supporting Information. Nevertheless, the values of electric field enhancement are still very significant and comparable in magnitude to alternative state-of-the-art approaches based on electric toroidal dipole resonances \cite{Yang:18}.

\section{Conclusions}

In this letter, we introduced a design principle to localize and enhance electric fields in high/moderate-index resonators, based on the flux conservation of the displacement current density.  We apply this design principle to a high-index ring resonator, showing that by squeezing its section through the introduction of a bottleneck, the electric field in correspondence of the bottleneck is increasingly enhanced as the width is reduced to keep the flux constant. Furthermore, due to the boundary conditions, the enhancement of the field is achieved in both the internal and external regions. Then, we also apply this principle to a moderate-index dielectric resonator, where the hypotheses behind the magnetoquasistatic approximation are only partially verified. Even though the achieved enhancement is reduced with respect to the previous ideal case, it is still very significant and comparable in magnitude to alternative state-of-the-art approaches based on electric toroidal dipole resonances. The introduced design principle, validated here in its simplest form, may constitute the first step into the engineering of the displacement current density, guiding the design of the topology and the geometry of complex dielectric structures to funnel light at will.

\begin{suppinfo}
The Supporting Information is available free of charge on the ACSPublications website at DOI:
In Fig S1 we carry out the comparison between the high-index ring with a bottleneck and the split-ring in the presence of a dipole excitation. In Fig. S2, we  investigate the spatial decay of the field enhancement in the external region (air) of a high-index ring with bottleneck as a function of the distance from the resonator's walls. In Fig. S3, the field enhancement in a Si ring resonator with a dipole excitation a function of its thickness, achieving field enhancement greater than $30 \times$. In Fig. S4, a Si ring resonator has been designed to operate in the near-infrared.   In Fig. S5 a silicon ring with a bottleneck is excited by a plane wave propagating in the plane of the ring.
	\end{suppinfo}

\providecommand{\latin}[1]{#1}
\makeatletter
\providecommand{\doi}
  {\begingroup\let\do\@makeother\dospecials
  \catcode`\{=1 \catcode`\}=2 \doi@aux}
\providecommand{\doi@aux}[1]{\endgroup\texttt{#1}}
\makeatother
\providecommand*\mcitethebibliography{\thebibliography}
\csname @ifundefined\endcsname{endmcitethebibliography}
  {\let\endmcitethebibliography\endthebibliography}{}


\end{document}